\documentclass[aps,twocolumn,prb,showpacs]{revtex4}% Physical Review B

\usepackage{graphicx}% Include figure files
\usepackage{dcolumn}% Align table columns on decimal point
\usepackage{bm}% bold math

\newcommand{\nt}{$\nu_T=1$}
\newcommand{\dl}{$d/\ell$}
\newcommand{\dlc}{$(d/\ell)_c$} 

\begin{document}

\title{Area dependence of interlayer tunneling in strongly correlated \\bilayer 2D electron systems at $\nu_{T}=1$}

\author{A.~D.~K. Finck$^1$, A.~R. Champagne$^1$, J.~P. Eisenstein$^1$, L.~N. Pfeiffer$^2$, and K.~W. West$^2$}
\affiliation{$^1$Condensed Matter Physics, California Institute of Technology, Pasadena CA 91125
\\
$^2$Bell Laboratories, Alcatel-Lucent, Murray Hill, NJ 07974}

\date{\today} 

\begin{abstract}
The area and perimeter dependence of the Josephson-like interlayer tunneling signature of the coherent \nt\ quantum Hall phase in bilayer two-dimensional electron systems is examined.  Electrostatic top gates of various sizes and shapes are used to locally define distinct \nt\ regions in the same sample.  Near the phase boundary with the incoherent \nt\ state at large layer separation, our results demonstrate that the tunneling conductance in the coherent phase is closely proportional to the total area of the tunneling region.  This implies that tunneling at $\nu_{T}=1$ is a bulk phenomenon in this regime.
\end{abstract}


\maketitle
Bilayer two-dimensional electron systems (2DESs) in a large perpendicular magnetic field $B$ support an unusual collective phase when the separation between the layers is sufficiently small, the temperature is sufficiently low, and the total density $n_T = n_1+n_2$ of electrons in the system equals the degeneracy $eB/h$ of a single spin-resolved Landau level created by the magnetic field\cite{perspectives}.  Interlayer and intralayer correlations are of comparable importance in this Landau level filling factor $\nu_T =\nu_1+\nu_2 =n_T/(eB/h)=1$ phase.  The system exhibits a number of striking physical properties, including Josephson-like interlayer tunneling\cite{spielman2000} and vanishing Hall and longitudinal resistances\cite{kellogg2004,tutuc2004,wiersma2004} when equal electrical currents are driven in opposition (counterflow) in the two layers.  The electronic correlations responsible for these properties may be understood in a number of equivalent languages, including that of itinerant ferromagnetism and that of excitonic Bose condensation\cite{jpe2004}.  Indeed, the remarkable transport properties observed in counterflow are very suggestive of excitonic superfluidity, although experiments have so far always detected small amounts of residual dissipation.

Interlayer tunneling provides particularly dramatic evidence that bilayer 2DESs at \nt\ are quite different at large and small interlayer separations.  When the effective layer separation (defined as the ratio \dl\ of the center-to-center quantum well spacing $d$ to the magnetic length $\ell=(\hbar/eB)^{1/2}$) is large, the interlayer tunneling conductance $dI/dV$ is strongly suppressed around zero interlayer voltage\cite{jpe92}.  This suppression reflects the strongly correlated nature of {\it single layer} 2DESs at high magnetic field.  In essence, an electron which is rapidly injected into a 2DES at $\nu <1$ is completely uncorrelated with the existing $N$-particle system and thus creates a highly excited $N+1$-particle state which can only slowly relax to equilibrium.  Similar considerations attend the rapid extraction of an electron from a 2DES at high magnetic field. Since interlayer correlations are negligible at large \dl, the net tunneling conductance spectrum of the bilayer is a simple convolution of the suppression effects in the individual layers.  In contrast, at sufficiently small \dl\ the interlayer tunneling conductance $dI/dV$ at \nt\ displays a strong and extremely sharp peak centered at $V =0$. The sharp peak in $dI/dV$ reflects a near-discontinuity in the tunneling $I-V$ curves at $V = 0$.  This is reminiscent of the dc Josephson effect, but it remains unclear how close the analogy is. 

The tunneling features observed at \nt\ and small \dl\ are widely believed to reflect the development of spontaneous interlayer quantum phase coherence among the electrons in the bilayer\cite{balents01,stern01,fogler01,joglekar01,fertig03,wang04,cooper04,wang05,dassarma06}. Electrons are just as strongly correlated with their neighbors in their own layer as they are with electrons in the opposite layer.  Within the exciton condensate language, every electron in the bilayer is bound to a correlation hole in the opposite layer and there is no energy cost associated with transferring an electron from one layer to the other. Crudely speaking, the Coulombic energy penalties associated with injection and extraction of electrons in the tunneling process are cancelled by the excitonic attraction of the electron to the hole in the final state.  

The purpose of the present paper is to report on an investigation of how interlayer tunneling in the coherent \nt\ phase is distributed across the sample area.  Via electrostatic gating we are able to study regions of different size and shape in the same sample.  For effective layer separations \dl\ not too much smaller than the critical value, approximately $(d/\ell)_c \approx 1.91$ as $T \rightarrow 0$ in our samples, we find that the zero bias tunneling conductance, $G(0) \equiv dI/dV$ at $V=0$, is essentially proportional to the sample area.  This suggests that tunneling at \nt\ is a bulk phenomenon in our samples, at least close to the critical layer separation.  In spite of the intuitive nature of this result, it is not obvious $a~priori$.  Indeed, the strong similarity between bilayer 2DESs at \nt\ and Josephson junctions suggests that tunneling currents may be restricted to micron-scale regions close to the source and drain contacts and not exist in the bulk of the sample\cite{stern01}. We discuss this and other theories of interlayer tunneling following the description of our experiment and its results.

The bilayer 2DES samples used in this experiment are GaAs/AlGaAs double quantum wells grown by molecular beam epitaxy (MBE). Two 18 nm GaAs quantum wells are separated by a 10 nm Al$_{0.9}$Ga$_{0.1}$As barrier.  Modulation doping with Si populates the ground subband of each well with a 2DES of nominal density $n_{1,2} \approx 5.5 \times 10^{10}$ cm$^{-2}$ and low temperature mobility $\mu \approx 10^6$ cm$^{-2}$/V s.  The tunnel splitting $\Delta_{SAS}$ between the lowest symmetric and antisymmetric states in the double well potential is estimated\cite{spielman2000,spielmanthesis} to be less than 100 $\mu$K; this is roughly six orders of magnitude smaller than the mean Coulomb energy $e^2/\epsilon \ell$ at \nt.  Independent electrical contact to the individual layers is achieved via a selective depletion method \cite{jpe90}, thus allowing the measurement of interlayer tunneling via standard lock-in techniques.

In order to minimize systematic errors in the tunneling conductances due to sample-to-sample variations, we employ a technique which allows multiple, independently controllable tunneling regions to be established on a single sample.  Two such samples, A and B, (both taken from the same parent MBE wafer) have been studied.  In both cases a large electrostatic gate is deposited on the back side of the sample.  This gate allows for global control of the electron density in the lower 2D electron gas in the active region of the sample. Smaller gates, of varying sizes, are deposited on the sample's top surface.  These gates allow for local control of the top 2DES density.  By adjusting the gates appropriately it is possible to establish a density balanced $\nu_T=1/2+1/2=1$ state underneath any one of the top gates while the remainder of the sample is in an unbalanced $\nu_T \neq 1$ configuration.  Since tunneling is strongly enhanced at \nt\ for \dl\ $<$ \dlc, but strongly suppressed\cite{jpe92} where $\nu_T \neq 1$, it is not difficult to separate out the conductance of the \nt\ region.  In fact, as we show below, this separation can also be done at zero magnetic field.  In sample A, the active region of the device is a 200 $\mu$m wide bar, fully underlaid by the large back gate.  This bar is crossed by four rectangular top gates, with lengths 100, 50, 20, and 10 $\mu$m. The top gates thus define four different regions with four different areas that are proportional to the top gate length.  In sample B the mesa bar has an inner section of width 100 $\mu$m and two outer sections of width 200 $\mu$m.  A 100 $\mu$m top gate crosses the inner section while two 50 $\mu$m top gates cross the outer sections. This sample allows the comparison of regions with the same area but different mesa widths and gate lengths.  Schematic diagrams of samples A and B are shown in Figs. 1a and 3a, respectively.

\begin{figure}
\includegraphics[width=3.25in, bb=71 155 384 468]{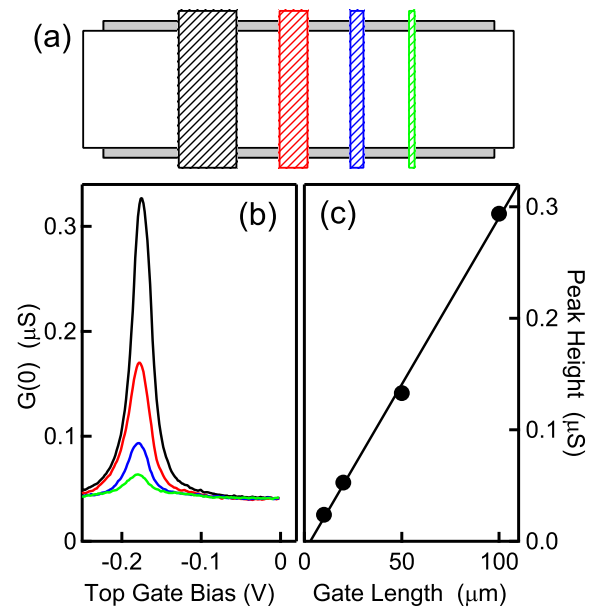}
\caption{\label{fig:area1}(color online) (a) Schematic diagram of sample A. White rectangle represents 200 $\mu$m-wide mesa containing the bilayer 2DES; ohmic contacts (not shown) to the individual 2DES layers are attached near both ends of the mesa.  Grey rectangle under the mesa is the global back gate; hatched rectangles crossing above the mesa are the various top gates under which tunneling is established. (b) Zero voltage tunneling conductance $G(0)$ versus top gate bias taken at $B=0$ and with an applied back gate bias of $-15.18$ V.  Peaks correspond, in order of decreasing height, to the 100, 50, 20, and 10 $\mu$m top gates. (c) Background-subtracted peak heights from (b) versus top gate length.  The line fitted to this data has an $x$-intercept of $3\pm 2$ $\mu$m.}
\end{figure}

%Area Tunneling sample:
At zero magnetic field momentum and energy conserving tunneling between parallel 2DES's can only occur when the subband energy levels in the two quantum wells line up\cite{jpe91}.  If the two 2DESs have equal electron densities this occurs at zero interlayer voltage, $V = 0$.  The width of the tunnel resonance is set by the lifetime of the electrons in the 2DESs; at low temperatures this is dominated by impurity scattering\cite{murphy95}. When the bilayer system is density imbalanced the tunnel resonance moves to a finite interlayer voltage proportional to the density difference.  We exploit this fact to allow for clean separation of the tunneling conductance under the top gates on our samples from the ``background'' tunneling coming from other portions of the device.  Figure 1b illustrates this with data from sample A.  For the data shown, the back gate bias has been set to -15.18 V; this lowers the density in the bottom 2DES and thereby imbalances the bilayer.  With the top gates set to zero bias a measurement of the net tunneling conductance at zero interlayer voltage, i.e. $G(0)$, yields a relatively small background value; the main resonance having been displaced to a finite interlayer voltage by the imbalance.  However, if the bias voltage on one of the top gates is swept, a resonance appears when the density of the top 2DES under that gate matches the density of the lower 2DES.  This is what Fig. 1b displays, there being one curve for each of the four top gates. Each gate produces a clear resonance in $G(0)$ when biased to about -0.17 V\cite{gateseparations}.  These resonances sit atop a background conductance of about 50 nS which arises from off-resonant tunneling from those regions of the device not under the swept top gate. 

Figure 1c shows the background-subtracted peak tunneling conductance for each of the four top gate traces shown in Fig. 1b.  The background subtraction is necessarily slightly different for each top gate. The tunneling conductance observed with all top gates grounded gives the amount of tunneling that comes from the entire tunneling region when imbalanced.  When a bias is applied to one of the top gates to create a balanced region underneath, the net area of the device which remains imbalanced is reduced by the area of the biased top gate.  We use this fact to calculate the correct background for each top gate.  The corrected peak heights $vs.$ top gate length are plotted in Fig. 1c.  The data are well-fitted by a straight line with a small $x$-intercept of approximately $3\pm2 ~\mu$m.  For the data shown in Fig. 1, $n_1 = n_2 = 3.3\times10^{10}$  cm$^{-2}$ at the resonance; but this same linear dependence on gate length is observed at all matched densities ranging from 2.9 to $3.4\times10^{10}$ cm$^{-2}$.  We therefore conclude that tunneling is proportional to area at $B=0$, as expected.  The positive $x$-intercept in the peak height versus top gate length relation may indicate that boundary effects are reducing the effective tunneling area.  For example, fringe fields along the edge of the gated regions could produce a density gradient over length scales comparable to the distance between the upper 2DES and the top gate (0.5 $\mu$m), thus reducing the area in the upper 2DES that has electron density matching that of the lower layer.

At \nt\ the situation is both simpler and more complex than at $B=0$.  The strong suppression\cite{jpe92} of tunneling in the unbalanced regions of the sample where $\nu_T \neq 1$ and the strong enhancement at small \dl\ of the tunneling in the balanced \nt\ regions under the top gates are both helpful in keeping the background tunneling conductance (typically $\sim$1 nS) a small fraction of the peak conductance at \nt\ (typically $\sim$10 - 1500 nS).  In contrast, it is much more difficult to ensure that the bilayer is density balanced at \nt\ than it is at zero magnetic field.  At \nt\ the tunnel resonance remains centered at zero interlayer voltage, regardless of the individual layer densities, provided $\nu_1+\nu_2 = \nu_T = 1$. Furthermore, the robustness of the \nt\ state against small antisymmetric shifts in the individual filling factors ($\nu_1 \rightarrow \nu_1 +\delta \nu$, $\nu_2 \rightarrow \nu_2 - \delta \nu$) complicates the determination that a balanced \nt\ system is present. In order to establish the gate voltages required for balanced \nt\ state underneath any given top gate, we first use an empirical calibration of the gates\cite{gatecalib} to choose approximate top and back gate voltages ($V_{top}$ and $V_{back}$) which will create $\nu_1 \approx \nu_2 \approx 1/2$ in each layer at some chosen magnetic field.  Since the empirical calibrations are reasonably good, this results in a clear tunneling peak at $V = 0$, thus proving that the system is at least close to the desired balanced \nt\ configuration.  Next, the zero bias tunneling conductance $G(0)$ is recorded as both the $V_{top}$ and $V_{back}$ are scanned across a window in voltage space.  Examination of the resulting data clearly reveals contours of constant total filling factor.  Small refinements of the magnetic field and the gate voltages are then made to first maximize $G(0)~vs.~B$ and then to perfect the symmetry\cite{symmetry} of the tunnel resonance shape ($dI/dV~vs.~V$).  

\begin{figure}
\includegraphics[width=3.25in, bb=78 74 384 302]{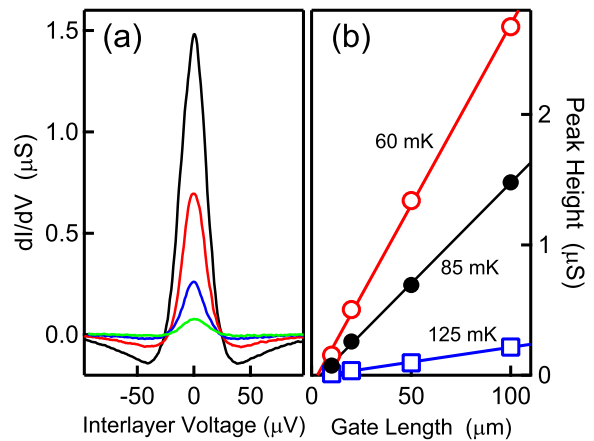} 
\caption{\label{fig:area2}(color online) Tunneling $vs.$ gate length at \nt\ in sample A at $d/\ell=1.81$. Gate layout as shown in Fig. 1. (a) Tunneling conductance resonances ($dI/dV~vs.~V$) at $T = 85$ mK. Gate lengths 100, 50, 20, and 10 $\mu$m; tallest to shortest.  (b) Peak heights plotted $vs.$ top gate length; open circles 60 mK, solid circles 85 mK, squares 125 mK. Straight lines are linear least squares fits.}
\end{figure}
Figure 2a shows typical tunneling conductance resonances ($dI/dV~vs.~V$) at \nt\ in sample A. For these data the temperature is $T=85$ mK and the effective layer spacing is $d/\ell=1.81$.  The four traces (tallest to shortest) correspond to tunneling under the four different top gates, with lengths 100, 50, 20, and 10 $\mu$m.  As explained above, there is negligible background tunneling coming from other parts of the sample.  Figure 2b shows the height of the tunneling peak at zero interlayer voltage, i.e. $G(0)$, $vs.$ top gate length at three temperatures, 60, 85, and 125 mK, for this same effective layer spacing.  The data reveal a clear linear dependence on gate length.  A small $x$-intercept, $5\pm 2~\mu$m is obtained from the linear least squares fits. Data very similar to that shown in Fig. 2 has been obtained at effective layer spacings ranging from $d/\ell=1.70$ to 1.88 and at temperatures from $T = 60$ mK to 300 mK.  In all cases the height of the tunneling peak scales linearly with gate length.

At smaller \dl\ deviations from the linear dependence on gate length begin to appear.  However, these deviations are most likely artifacts arising from the reduced sheet conductivity of the 2D systems at these lower densities and the concomitant enhanced tunneling conductance.  When the sheet conductivity of the 2DESs no longer greatly exceeds the tunneling conductance, voltage drops develop within the 2D layers.  The resonance in $dI/dV$ becomes broadened and suppressed in height.  Since this effect is more pronounced in larger tunneling areas than in smaller ones, the tunneling peak height becomes sub-linear in gate length.  For this reason we restrict our attention here to effective layer spacings and temperatures relatively close to the phase boundary\cite{champagne08} separating the interlayer coherent and incoherent states at \nt. In this regime we are confident that sheet resistance effects are negligible.

The data in Fig. 2 strongly suggest that the interlayer tunneling conductance in the coherent \nt\ state is proportional to the area of the tunneling region.  However, since in sample A it is the length of the gated regions that is varied while the width (200 $\mu$m) is constant, it remains possible that the tunneling conductance is simply proportional to gate length, not area.  For this reason we now turn to the results obtained from sample B.  This device, illustrated schematically in Fig. 3a, contains three top-gated regions; two are 200 $\mu$m wide by 50 $\mu$m long rectangles and the third is a 100$\times$100 $\mu$m square.  These three regions all have the same area but differ by a factor of two in gate length and width.  Since the two rectangular gates are identical in size and shape, we focus here on the comparison of the inner square top gate and one of these outer rectangular gates.  Direct comparisons of tunneling in the two rectangles justify this simplification.

\begin{figure}
\includegraphics[width=3.35in, bb=71 155 394 465]{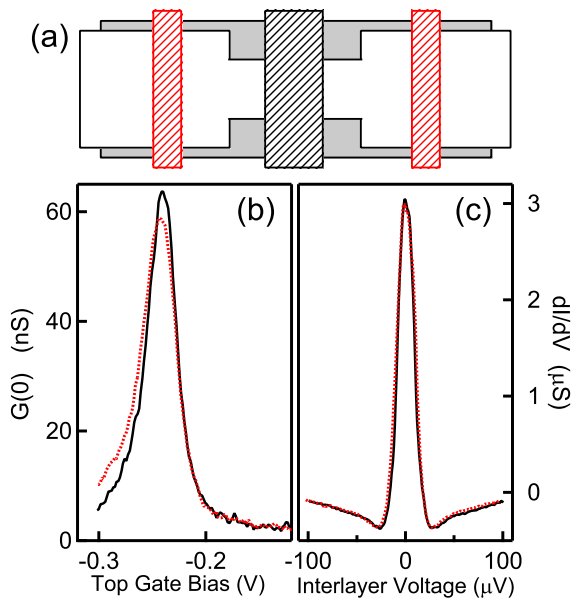} 
\caption{\label{fig:perimeter}(color online) (a) Schematic diagram of the mesa and gates in the active region of sample B. (b) Zero voltage tunneling conductance $G(0)$ versus top gate bias taken at $B=0$ and with an applied back gate bias of $-17.08$ V. (c) Tunneling conductance resonances ($dI/dV~vs.~V$) at \nt\ with $d/\ell$ = 1.64 and $T = 61$ mK.  In (b) and (c) the solid trace (black) corresponds to the 100 $\mu$m square tunneling region and dotted trace (red) to the $200\times50~\mu$m rectangular region.  The red trace in (c) has been multiplied by 1.086, as described in the text.}
\end{figure}
 
Figure 3b shows typical tunneling conductance resonances in sample B at zero magnetic field. As with sample A, at $B=0$ it is most informative to plot $G(0)$, the tunneling conductance at zero interlayer voltage, versus top gate bias at fixed back gate bias, here $V_{back}=-17.08$ V.  Although sample B was taken from the same MBE wafer as sample A, its tunneling conductance per unit area is about a factor of $\sim2$ smaller.  We attribute this difference to known wafer-scale variations in the thicknesses of the various epilayers in the sample\cite{estimate}. More significantly, Fig. 3b shows that the peak tunneling conductance of the 100 $\mu$m square region is about 10 percent larger, and the resonance width about 10 percent smaller, than that in the $200\times50~\mu$m rectangle. While these differences might be due to differing degrees of density inhomogeneity in the two regions, their exact origin is unknown.  In the following, when comparing the tunneling conductances at \nt\ in the square and rectangular regions of sample B, we correct for this small effect by multiplying the tunneling data from the rectangular region by the appropriate factor deduced from the zero field data.  Figure 3c illustrates the results of this comparison of tunneling conductance spectra ($dI/dV~vs.~V$) at \nt\ for $d/\ell = 1.64$ and $T=60$ mK; there is essentially no difference in the spectra from these two differently shaped regions. Very similar results have been found at densities corresponding to $1.60 \leq d/\ell \leq 1.79$ and temperatures from $T=60$ to 300 mK.

Taken together, the data presented in Figs. 2 and 3 provide strong evidence that the enhanced tunneling conductance characteristic of the coherent \nt\ bilayer state is proportional to the area of the tunneling region, at least over a limited range of \dl\ and temperature near the phase boundary.  The possibility that the conductance is instead proportional to the perimeter of the tunneling region, or simply the length of that region along the mesa boundary is not consistent with the present findings.

The intuitive character of our findings belies the subtle nature of tunneling in the coherent bilayer \nt\ state.  The long-wavelength effective Hamiltonian for this many-electron state is closely analogous to that of a Josephson junction. As pointed out by Stern, $et~al.$\cite{stern01} this analogy leads to a sine-Gordon equation for the collective phase variable $\phi$ representing the orientation of the pseudospin magnetization in the ferromagnetically ordered state\cite{yang94,moon95}. In an ideal, disorder-free sample this equation leads to the conclusion that interlayer tunnel currents (proportional to sin$\phi$) are restricted to narrow regions near where the current enters and leaves the bilayer system\cite{sourcedrain}.  The width $\lambda_J$ of these regions is determined by the pseudospin stiffness and the tunnel splitting $\Delta_{SAS}$, and is estimated to be in the few $\mu$m range.  In contrast to the results presented here, this scenario would not result in a tunneling conductance proportional to area\cite{perimeter}.

More realistic models of the distribution of tunneling in the coherent \nt\ state have recently been advanced by Fertig and Murthy\cite{fertig05} and by Rossi, {\it et al.}\cite{rossi05}. Both models attempt to incorporate the crucial effects of disorder on the tunneling.  In Fertig and Murthy's picture\cite{fertig05} fluctuations in the 2DES density lead to a complex network of channels and nodes, covering the entire sample, in which the coherent \nt\ state exists. Tunneling occurs at the nodes of the network.  Since the disorder length scale is expected to be small ($<1~\mu$m, set by the separation between the dopant layers and the 2DESs), the number of nodes, and therefore the net tunneling conductance, is predicted to be proportional to sample area in large devices, in agreement with our results.  In the model of Rossi, {\it et al.}\cite{rossi05} disorder along the physical edge of the sample is expected to dominate when the coherent phase is well-developed (i.e. at small \dl\ and very low temperature).  In this regime they predict that the tunneling conductance will be proportional to the length along the sample edge, not the total area.  Although this is in conflict with our observations, we stress that our results are confined to relatively large \dl, in fairly close proximity to the phase boundary.

In summary, we have found strong evidence that interlayer tunneling in the coherent bilayer \nt\ phase near the critical layer separation is proportional to the area of the tunneling region, with some evidence for small edge corrections.  We therefore conclude that tunneling in the coherent \nt\ state, much like tunneling at zero magnetic field, is a bulk phenomenon in our samples.

We are pleased to thank Herb Fertig and Allan MacDonald for enlightening discussions.  This work was supported via NSF Grant No. DMR-0552270 and DOE Grant No. DE-FG03-99ER45766.


\begin{references}

\bibitem{perspectives} For an early review, see the chapters by S.M. Girvin and
A.H. MacDonald, and by J.P. Eisenstein in \emph{Perspectives in Quantum Hall Effects}, edited by S. Das Sarma and A. Pinczuk, (John Wiley, New York, 1997).

\bibitem{spielman2000} I.B. Spielman {\it et al.}, Phys. Rev. Lett. {\bf 84}, 5808 (2000) and Phys. Rev. Lett. {\bf 87}, 036803 (2001).

\bibitem{kellogg2004} M. Kellogg {\it et al.}, Phys. Rev. Lett. {\bf 93}, 036801 (2004).

\bibitem{tutuc2004} E. Tutuc, M. Shayegan, and D. Huse, Phys. Rev. Lett. {\bf 93}, 036802 (2004).

\bibitem{wiersma2004} R. Wiersma {\it et al.}, Phys. Rev. Lett. {\bf 93}, 266805 (2004).

\bibitem{jpe2004} J.P. Eisenstein and A.H. MacDonald, Nature {\bf432}, 691 (2004).

\bibitem{jpe92} J.P. Eisenstein, L.N. Pfeiffer, and K.W. West, Phys. Rev. Lett. {\bf 69}, 3804 (1992).

\bibitem{balents01} L. Balents and L. Radzihovsky, Phys. Rev. Lett. {\bf 86}, 1825 (2001).

\bibitem{stern01} A. Stern {\it et al.}, Phys. Rev. Lett. {\bf 86}, 1829 (2001).

\bibitem{fogler01} M. M. Fogler and F. Wilczek, Phys. Rev. Lett. {\bf 86}, 1833 (2001).
\bibitem{joglekar01} Y. N. Joglekar and A. H. MacDonald, Phys. Rev. Lett. {\bf 87}, 196802 (2001).

\bibitem{fertig03} H. A. Fertig and J. P. Straley, Phys. Rev. Lett. {\bf 91}, 046806 (2003).

\bibitem{wang04} Z. Q. Wang, Phys. Rev. Lett. {\bf 92}, 136803 (2004).

\bibitem{cooper04} R. L. Jack, D. K. K. Lee, and N. R. Cooper, Phys. Rev. Lett. {\bf 93}, 126803 (2004) and Phys. Rev. B {\bf 71}, 085310 (2005).

\bibitem{wang05} Z. Q. Wang, Phys. Rev. Lett. {\bf 94}, 176804 (2005).

\bibitem{dassarma06} K. Park and S. Das Sarma, Phys. Rev. B {\bf 74}, 035338 (2006).

\bibitem{yang94} Kun Yang {\it et al.}, Phys. Rev. Lett. {\bf 72}, 732 (1994).

\bibitem{moon95} K. Moon, {\it et al.}, Phys. Rev. B {\bf 51}, 5138 (1995).

\bibitem{spielmanthesis} I.B. Spielman, Ph.D. thesis, Caltech(2004).

\bibitem{jpe90} J.P. Eisenstein, L.N. Pfeiffer, and K.W. West, Appl. Phys. Lett. {\bf 57}, 2324 (1990).

\bibitem{jpe91} J.P. Eisenstein, L.N. Pfeiffer, and K.W. West, Appl. Phys. Lett. {\bf 58}, 1497 (1991).

\bibitem{murphy95} S.Q. Murphy, J.P. Eisenstein, L.N. Pfeiffer, and K.W. West, Phys. Rev. B {\bf 52}, 14825 (1995).

\bibitem{gateseparations} The top gates lie approximately 0.5 $\mu$m above the bilayer 2DES, while the back gate is roughly 50 $\mu$m below them.

\bibitem{gatecalib} The back gate is calibrated by observing Shubnikov-de Haas oscillation in the resistivity of the lower 2DES.  The top gates are then calibrated relative to the back gate by observing the tunneling resonances they induce at zero magnetic field; see Fig. 1.

\bibitem{symmetry} We find that there is a weak asymmetry of the tunneling resonance shape at \nt\ when the bilayer is imbalanced.  The origin of this asymmetry is not understood, but it may result from residual incoherent tunneling for which imbalance-induced asymmetry is both expected and observed. 

\bibitem{champagne08} A.R. Champagne, J.P. Eisenstein, L.N. Pfeiffer and K.W. West, Phys. Rev. Lett. {\bf 100}, 096801 (2008). 

\bibitem{estimate} A simple estimate shows that only a $\sim 2$ percent increase in the nominal 10 nm thickness of the barrier layer is needed to reduce the tunneling conductance at $B=0$ by a factor of 2. 

\bibitem{sourcedrain} In our experiments current in one 2DES layer is fed into the tunneling region from one side (say from the left in Fig. 1a) and then removed from that same side but in the other 2DES layer.

\bibitem{perimeter} Such a model would presumably predict that the tunneling conductance would be proportional to the width or perimeter of the tunneling region.  Neither possibility is consistent with our data.

\bibitem{fertig05} H.A. Fertig and G. Murthy, Phys. Rev. Lett. {\bf 95}, 156802 (2005).

\bibitem{rossi05} E. Rossi, A. Nunez, and A.H. MacDonald, Phys. Rev. Lett. {\bf 95}, 266804 (2005).

\end{references}
\end{document}